\documentstyle[11pt]{article}
\def\thefootnote{\fnsymbol{footnote}}

\newcommand{\nc}{\newcommand}
\nc{\be}{\begin{equation}}
\nc{\ee}{\end{equation}}
\nc{\bea}{\begin{eqnarray}}
\nc{\eea}{\end{eqnarray}}
\nc{\rbo}{\raisebox}
\nc{\cH}{{\cal H}}
\nc{\um}{\frac12}
\nc{\Rr} {\rangle}
\nc{\Ll} {\langle}
\nc{\rmi}[1]{{\mbox{\small #1}}}
\nc{\eq}{Eq.~}
\nc{\nr}[1]{(\ref{#1})}
\nc{\ul}{\underline}
\nc{\cM}{{\cal M}}
\nc{\mc}{\multicolumn}
\begin{document}
\begin{titlepage}
\begin{flushright}
CERN-TH 7291/94 \\
DFTT 22/94\\
DF-UNICAL/94-15\\
MS-TPI-94-6 \\
July 1994
\end{flushright}
\vskip0.5cm
\begin{center}
{\Large\bf  Rough Interfaces Beyond the Gaussian Approximation}
\vskip0.2cm
\end{center}
\vskip 0.6cm
\centerline{M. Caselle$^a$, R. Fiore$^b$, F. Gliozzi$^a$,}
\centerline{M. Hasenbusch$^c$, K. Pinn$^d$ and S. Vinti$^a$}
\vskip 0.6cm
\centerline{\sl  $^a$ Dipartimento di Fisica
Teorica dell'Universit\`a di Torino}
\centerline{\sl Istituto Nazionale di Fisica Nucleare, Sezione di Torino}
\centerline{\sl via P.Giuria 1, I--10125 Torino, Italy
\footnote{e--mail: caselle, gliozzi, vinti~@to.infn.it}}
\vskip .2 cm
\centerline{\sl $^b$ Dipartimento di Fisica, Universit\`a della Calabria}
\centerline{\sl Istituto Nazionale di Fisica Nucleare, Gruppo collegato
di Cosenza}
\centerline{\sl Rende, I--87030 Cosenza, Italy
\footnote{e--mail: fiore~@fis.unical.it}}
\vskip .2 cm
\centerline{\sl  $^c$ Theory Division, CERN}
\centerline{\sl CH--1211 Geneva 23, Switzerland
\footnote{e--mail: hasenbus~@surya11.cern.ch}}
\vskip .2 cm
\centerline{\sl $^d$ Institut f\"ur Theoretische Physik I,
Universit\"at M\"unster}
\centerline{\sl Wilhelm--Klemm--Str. 9, D--48149 M\"unster, Germany
\footnote{e--mail: pinn~@xtp139.uni-muenster.de}}
\vskip 0.6cm

\begin{abstract}
\vskip0.2cm
We compare predictions of the Capillary Wave Model
beyond its Gaussian approximation with Monte Carlo
results for the energy gap and the surface energy
of the 3D Ising model in the scaling region.
Our study reveals that the finite size effects of these
quantities are well described by the Capillary Wave
Model, expanded to two--loop order (one order beyond
the Gaussian approximation).
\end{abstract}
\end{titlepage}

\setcounter{footnote}{0}
\def\thefootnote{\arabic{footnote}}

\section{Introduction}

Soft modes play an essential role in the description of
{\em  finite size effects} (FSEs) in the fluid interface's free energy
(see for instance \cite{GFP} and references therein).
3D spin systems offer a simple context where these effects appear
and can be studied, e.g.\ by using numerical simulations
to check theoretical predictions.
It is  well known that, between the roughening and the critical
temperature of the 3D Ising model, interfaces are dominated
by long wavelength fluctuations (i.e.\ they behave as
{\em  fluid} interfaces).
Because of these fluctuations, a complete control on the description
of interfaces,
starting from the microscopic Hamiltonian, has not yet been reached
in the rough phase. The usual approach consists in assuming an
effective Hamiltonian describing the collective degrees of freedom.

An effective model widely used to describe a
rough interface is the {\em  capillary wave model} (CWM) \cite{BLS}.
In its simplest formulation one assumes an effective Hamiltonian
proportional to the variation of the surface's area with respect to the
classical solution. Because of its non--polynomial nature, until recently 
the CWM has been studied only in its quadratic approximation, the
Hamiltonian being equivalent to a (massless) 2D Gaussian model.
The CWM has been often identified with the Gaussian model.
This model was shown to give a good description
of several features of the interfaces of the 3D Ising model, like the
logarithmic growth (as a function of the lattice size)
of the interfacial width in the whole rough phase~\cite{HeP}, and the
FSEs (as a function of the shape of the lattice)
of the free interface energy in the scaling region of the
model~\cite{cgv}.

The corrections that arise when passing from the classical
approximation to the Gaussian one
\cite{IZ,BB,cgv}, (which we shall term from now on
{\em  one--loop} contributions),
only depend on one adimensional parameter
(namely on the asymmetry $u=L_2/L_1$ of the transverse sizes of the
lattice) and on the boundary conditions.
They reduce on symmetric lattices ($u=1$) to the well known finite--size
behaviour of the free energy \cite{FPBZ,GM} in the large $L$ limit
\be
 \frac{F}{k_B T} \propto \sigma L^2~,
\label{class}
\ee
$\sigma$ being the reduced interface tension.

As mentioned above, the CWM has been often identified with  its Gaussian
approximation.  However, the full CWM is interesting in itself because
of its simple geometrical meaning (see for instance Ref.~\cite{Zia} 
and references therein). Also, it coincides with the
Nambu--Goto string action in a particular gauge (see the next section).
The improvement of Monte Carlo simulations reached in the
last years allows now to test the CWM beyond the Gaussian approximation.
This is the aim of this paper, in which we investigate the presence  of
corrections to the pure Gaussian description of the free energy of a
rough   interface in the scaling region of the 3D Ising model.

A similar analysis was recently  made in the
context of the 3D three--state Potts  model \cite{PV}. It was
shown  that higher  order corrections to the Gaussian approximation of
the CWM (from now on called {\em two--loop} contributions) give
contributions to the  functional form of the interface's free energy
which can be exactly estimated.

The contributions to the interface free energy (i.e.\ vacuum
diagrams on finite volumes) due to higher order expansion of
the CWM Hamiltonian are finite and can be evaluated in a simple way
\cite{DF}.
In this paper we present evidence that
the results are independent of the regularization used.

It turns out that the two--loop contributions do not  depend only on the
asymmetry parameter $u$ but also on the adimensional expansion parameter
proportional to the minimal  area of the surface, namely $\sigma
A\equiv\sigma L_1 L_2$. As a consequence and in contrast to what happens
for the Gaussian model, even for symmetric ($L_1=L_2\equiv L$) lattices
the  finite--size behaviour of the free energy defined in Eq.~(\ref{class})
gets corrections proportional to $(\sigma L^2)^{-1}$  which are
important in extracting the interface tension value using a fitting
procedure. In this paper we discuss this picture for the Ising model,
checking the theoretical prediction by means of Monte Carlo simulations
and using very different techniques and algorithms.

Finally, let us remark that
the 3D Ising model is related through
duality to 3D $Z_2$ gauge theory.
 In particular the physics of
interfaces is directly linked to the physics of their dual gauge
observables, i.e.\ the Wilson  loops, and, as a consequence, to the
properties of the  chromo--electric flux tube in the confining phase
\cite{cfggv}. This means that all the results that we describe in
this paper have a direct counterpart in the context of Lattice Gauge
Theories, and  could help to better understand the  possible string--like
descriptions  of their infrared behaviour.

\vskip0.3cm
The paper is organized as follows: in Sec.~2 we review the model for
rough interfaces near the continuum. In Sec.~3 we present an
alternative derivation of the two--loop calculation given
in Ref.~\cite{DF}.
In Sec.~4 we introduce and discuss the different observables
and Monte Carlo techniques used to
test the CWM. In Sec.~5 we analyze the numerical results.

\section{The Models}

\subsection{The Ising Model}
We consider the 3D Ising model on a regular cubic lattice
of size $L_1$, $L_2$ in the $x$--,
$y$--directions, where $L_2 \geq L_1$,
and size $t$ in the $z$--direction. In the $x$-- and the $y$--direction
periodic boundary conditions are imposed,
 while in the $z$--direction  either
periodic or anti-periodic boundary conditions are used,
 depending on the Monte Carlo
method that is applied.
The Hamiltonian is defined by
\begin{equation}
  H = - \sum_{<ij>} s_i s_j ~~,
\label{IsH}
\end{equation}
where the sum is over all nearest--neighbour pairs
$<ij>$, and $s_i=\pm1$.
The corresponding partition function is
\begin{equation}
Z_I=\sum_{\left\{s_i=\pm1\right\}} e^{-\beta H}~~,
\end{equation}
where $\beta=1/\left(k_B T\right)$.
The critical temperature of the model is estimated to be
$\beta_c=0.221652(3)$ \cite{betac}. The most precise
estimate for the roughening coupling is $\beta_r= 0.4074(3)$ \cite{MH}.

We study the model between the critical and the roughening
temperature, namely in the region $\beta_r>\beta>\beta_c$.

While in infinite volume, for $\beta>\beta_c$, the system shows a
spontaneous symmetry breaking, in finite volume this cannot occur, and
interfaces appear, separating extended domains of different
magnetization.

In particular, we will  consider  interfaces
that are parallel
to the $x-y$ plane and fluctuate freely in the third,
orthogonal, $z$--direction. In section 4 we shall discuss
how such interfaces can be generated
in a Monte Carlo simulation.

Above the roughening temperature ($\beta_r>\beta>\beta_c$) the step
free energy of the interface goes to zero. As a consequence
the interface behaves essentially as a 2D critical system: it may be
freely translated through the medium and the long-wavelength,
transverse fluctuations in the interface position, i.e.\ capillary waves,
have  a small cost in energy (hence cannot be neglected in
calculations).
They can be viewed as the Goldstone modes associated with the
spontaneous breaking of the transverse translational invariance
\cite{ML}.
To describe the interface free energy one is therefore forced to assume
an effective model. On the other side, one has the advantage of
choosing a Hamiltonian defined directly on the continuum to make
analytical computations.

\subsection{The Effective Model}
We assume  the effective Hamiltonian of the interface to be
proportional to its area. Denoting by $x_i(\xi_1,\xi_2)$, $i=1,2,3$,
the coordinates of a point of the interface as functions of the
parameters $\xi_\alpha$, $\alpha=1,2$, with $0\leq\xi_\alpha\leq1$, we
can write the area in the standard reparametrization invariant form,
\be
{\cal A}=\int_0^1 d\xi_1 \int_0^1 d\xi_2~ \sqrt{g}~~,
\label{area}
\ee
where $g={\rm  det}(g_{\alpha\beta})$, with
\be
g_{\alpha\beta}=\frac{\partial x_i}{\partial \xi_\alpha}
\frac{\partial x^i}{\partial\xi_\beta}~~.
\ee
A system with a Hamiltonian proportional to
${\cal A}$ coincides with the Nambu--Goto model for the
bosonic string. The corresponding quantum theory  is anomalous:
depending on the quantization method
one finds either the breaking of rotational invariance  or the
appearance of  interacting longitudinal modes (Liouville field). We are,
however, interested in interfaces of very large size  where these
difficulties disappear and the rotational invariance is restored
\cite{olesen}; in the infrared limit the theory flows to the massless
Gaussian model. In order to study the first perturbative correction to
this limit it is convenient to assume that the main contribution to
the interface free energy is given in this region by small and smooth
deformations of the minimal surface (which is a flat torus of area
$L_1L_2$). More  precisely, we assume that there are no foldings nor
self-intersections nor overhangs\footnote{
{}From the microscopic point of view this is not
obvious {\em  a priori}: in fact it has been observed that the interface
at small scales is much more similar to a sponge than to a smooth
surface \cite{DHMMPW,cgv2}. The strong linear correlation
between the area and the
genus of the surface (number of microscopic handles), together with
the evidence that the partition function summed over all genera
behaves like a smooth surface, has led to conjecture that a simple
non--perturbative renormalization of the physical quantities
associated to the surface occurs \cite{cgv2}.}.
Under these conditions we can choose as parameters the
two longitudinal coordinates $x=x_1$ and $y=x_2$, by putting
$\xi_1=x_1/L_1$ and $\xi_2=x_2/L_2$; as a consequence,
 the transverse displacement $x_3$ of the surface
becomes  a single--valued function of them: $x_3=\phi(x,y)$. In such a
frame Eq.~(\ref{area}) can be written as
\begin{equation}
A\left[ \phi\right]= \int_0^{L_1} dx \int_0^{L_2}dy~
 \sqrt{1+\left(\frac{\partial \phi}{\partial x}\right)^2
+\left(\frac{\partial \phi}{\partial y}\right)^2}~~.
\end{equation}
\def\h{{\cal H}}
Using these notations,  the free energy of a fluid interface can be
described by
\begin{eqnarray}
F &=& -~ k_B T~ \ln Z\\
Z&=& \lambda~e^{-\sigma L_1 L_2} Z_q\left(\sigma, L_1,L_2\right)
\label{zeta}\\
Z_q&=& \int \left[ D\phi\right]\exp\left\{- \h\left[
\phi\right]\right\}~~,
\label{zetaq}
\end{eqnarray}
where $\lambda$ is an undetermined constant within this approach.
The  reduced Hamiltonian $\h$ is given by
\begin{equation}
\h\left[ \phi\right]=\sigma \left(A\left[\phi\right]-L_1L_2\right)~~.
\label{cwm}
\end{equation}
In words,  $\h[\phi]$ is given by the change produced by the
deformation $\phi$ in the interface's area, measured in units of the
interface tension $\sigma$. \footnote{
Far from the scaling region, the interface tension 
$\sigma$ in Eq.~(\ref{cwm})
should be replaced by the stiffness $K$.
However, in the
scaling region, as the bulk correlation length increases and the
rotational invariance is restored, $K$ approaches $\sigma$ and for all
the values of $\beta$ that we studied the difference between them can be
safely neglected \cite{HeP}.}

One can then take into account the quantum contributions by
expanding Eq.~(\ref{cwm}) in the natural adimensional parameter
$ (\sigma A)^{-1}$, $A=L_1L_2$. The interface tension is the
only dimensional parameter of this theory.

Coming back to  the adimensional parameters $\xi_\alpha=x_\alpha/
L_\alpha$, and putting $\phi'=\sqrt{\sigma}\phi$, the Hamiltonian
can  be written as
\begin{eqnarray}
\h\left[ \phi\right]&=&\sigma A \int_0^1 d\xi_1 \int_0^1 d\xi_2~
\left[ \sqrt{1+\frac{1}{\sigma A}(\nabla\phi)^2}-1\right]
\label{cwmham}\\
(\nabla\phi)^2&=&
u\left(\frac{\partial \phi}{\partial \xi_1}\right)^2
+\frac{1}{u}\left(\frac{\partial \phi}{\partial \xi_2}\right)^2~~,
\end{eqnarray}
where $u= L_2/L_1$, and the primes have been omitted.
For $(\sigma A)^{-1}\rightarrow 0$, expanding up to the second order,
one obtains
\begin{eqnarray}
\h\left[ \phi\right]&=& \h_g\left[ \phi\right] -\frac{1}{8\sigma A}
\h_{p}\left[ \phi\right] + O\Bigl((\sigma A)^{-2}\Bigr) \label{Atot}\\
\h_g\left[ \phi\right]&=&  \frac{1}{2} \int_0^1 d\xi_1 \int_0^1 d\xi_2~
(\nabla\phi)^2 \label{Ag}\\
\h_{p}\left[ \phi\right]&=&
\int_0^1 d\xi_1 \int_0^1 d\xi_2~ \Bigl((\nabla\phi)^2\Bigr)^2
\label{Ap}~~.
\end{eqnarray}
Retaining only the quadratic term of Eq.~(\ref{Ag}) in the Hamiltonian 
(\ref{Atot}), one obtains from Eq.~(\ref{zetaq}) the
one--loop contribution which constitutes the Gaussian approximation
\begin{equation}
Z_q^{(g)} \left(u \right) = \frac{1}{\sqrt{u}}
\Bigl| \eta\left(i u\right)/\eta\left(i\right) \Bigr|^{-2}~~,
\label{Zg}
\end{equation}
where $\eta$ is the Dedekind eta function
\begin{equation}
\eta(\tau)=q^{1/24}\prod_{n=1}^{\infty}\left(1-q^n\right)~~,
\quad\quad q\equiv \exp(2\pi i \tau)~~.
\label{eta}
\end{equation}
This is a well known result in string theory and conformal field
theory, and coincides\footnote{The constant $\eta\left(i \right)$
has been introduced just to normalize $Z_q^{(g)}\left(1\right)=1$.}
with the partition function of a 2D conformal invariant free
boson on a torus of modular parameter $\tau=i u $ \cite{IZ,BB,cgv}.

Within this approximation, the interface partition function takes the
form
\begin{equation}
Z~ =~ \lambda~ e^{-\sigma L_1 L_2} Z_q^{(g)}\left(u\right)~~.
\label{1loop}
\end{equation}
For $u=1$, i.e.\ $L_1=L_2\equiv L$, one recovers the well
known finite size behaviour of the interface partition function given in
Eq.~(\ref{class}),
\begin{equation}
Z~ =~ \lambda~ e^{-\sigma L^2}~~.
\label{class2}
\end{equation}
It has been already verified \cite{cgv} that the inclusion of the
one--loop contribution (\ref{1loop}) allows to 
describe accurately finite size effects on asymmetric lattices ($u>1$):
these turn out to be strong enough to make the classical
approximation (\ref{class2}) completely inadequate. 
However, this result does not give a definitive answer about the
reliability of the CWM hypothesis expressed by Eq.~(\ref{cwm}): it is 
in fact well known that the Gaussian model is the fixed point of a wide
class of possible effective descriptions \cite{ML}.
To identify the particular effective Hamiltonian which describes the
free fluid interface it is then crucial to test higher--order
contributions.

Let us also stress that, among various possibilities, Eq.~(\ref{cwm}) 
is the simplest and most intuitive from a geometrical point of view.
Moreover, it does not add any new free parameter and,
even if this hypothesis is rather old \cite{BLS}, it has
never been tested beyond the Gaussian approximation until
recently \cite{PV}.

\section {Two-Loop Calculation}
In this section we calculate the  contribution to the partition
function $Z$  from the term given in the  Eq.~(\ref{Ap}) which is
the first correction to the Gaussian Hamiltonian.
We write $Z$ in the form
\begin{eqnarray}
Z &=& \lambda_o~{\rm  e}^{-\sigma_o A}Z_q^{(g)}\left(u\right)
Z_q^{(2l)}\left(u,\sigma_o A,\varrho_o\right)
\label{zeta2l}\\
Z_q^{(2l)} &=& 1+\frac{\varrho_o}{8\sigma_oA\,Z_q^{(g)}}
\int \left[ D\phi\right]
\h_p\left[ \phi\right]e^{- \h_g\left[ \phi\right]}
+ O\left( (\sigma A)^{-2}\right)\nonumber\\
&=&1+\frac{\varrho_o}{8\sigma_oA}\Ll\int_0^1 d\xi_1 \int_0^1 d\xi_2~
\Bigl((\nabla\phi)^2\Bigr)^2\Rr~~,
\label{zetaq2l}
\end{eqnarray}
where the expectation value is taken in the free theory.
In this section the subscript $~_o$ is added to the bare quantities in
order to distinguish them from the corresponding renormalized ones.
In the Nambu-Goto  model $\varrho_o=1$.

Eq.~(\ref{zetaq2l}) shows that the two--loop contribution can be 
expressed
in terms of products of double derivatives of the free Green function.
Wick's theorem gives
\bea
\Ll\int_0^1 d\xi_1 \int_0^1 d\xi_2~
\Bigl((\nabla\phi)^2\Bigr)^2\Rr&=&\Bigl[3u^2(\partial^2_{\xi_1}G)^2+
3u^{-2}(\partial^2_{\xi_2}G)^2+
\nonumber\\
2\partial^2_{\xi_1}G
\partial^2_{\xi_2}G&+&
4(\partial_{\xi_1}\partial_{\xi_2}G)^2\Bigr]_{\xi\to\xi'}
\label{limit}
\eea
with
\be
G(z-z')=\Ll\phi(\xi)\phi(\xi')\Rr~~,
\ee
where we have introduced the complex variable $z=\xi_1+iu\xi_2$.

Let us denote by $\{\omega\}$ the period lattice,
namely the set of points of the complex plane of the form
$\{\omega={\rm  m}+i{\rm  n}u,~{\rm  m,n}\in Z\} $~.
Then the periodic boundary conditions can be written simply as
\be
G(z+\omega)=G(z)~~.
\label{bc}
\ee
The Green function should also satisfy
\be
-\Delta\;\Ll\phi(\xi)\phi(0)\Rr=\delta^{(2)}(\xi)-1
\label{de}
\ee
where
\be
\Delta=u\frac{\partial^2}{\partial\xi_1^2}+\frac1u
\frac{\partial^2}{\partial\xi_2^2}~~.
\ee
The $-1$ term on the right-hand side of Eq.~(\ref{de}) represents the
subtraction of the zero mode due to the translational invariance of
the interface in the $z$--direction: this is the standard procedure
making  $\Delta$ invertible in the subspace orthogonal to the zero
mode (see for instance Ref.~\cite{ID}).

The solution of the above equation can be expressed
in terms of the Weierstrass $\sigma$ function, defined through the
infinite product
\be
\sigma(z)=z\prod_{\omega\not=0}\left(1-\frac{z}{\omega}\right)
{\rm  e}^{z/\omega+\um(z/\omega)^2}~~.
\ee
We can write the solution as a sum of three terms
\be
G(z)=-\frac{1}{2\pi}\ln\Bigl|\sigma(z)\Bigl|~
+~\frac{\pi E_2(iu)}{12}~\Re e(z^2)+\frac{1}{2u}\Bigl(\Im m(z)\Bigr)^2~~,
\label{sol}
\ee
where $E_2(\tau)$ is the first Eisenstein series
\begin{equation}
E_2(\tau)=1-24\sum_{n=1}^{\infty}\frac{n~ q^n}{1-q^n}~~,
\quad\quad q\equiv \exp(2\pi i \tau)~~.
\label{E2}
\end{equation}
The last term in Eq.~(\ref{sol}) accounts for the zero mode subtraction,
while the first two may be thought of as the real part of an analytic
function $f(z)$, hence they satisfy $\Delta\Re e f(z)=0$ outside the
singularities at $z\in \{\omega\}$. For $z$ near a node of the period
lattice the second term is regular while the first one behaves like
$-\frac{1}{ 2\pi}\ln\vert z-\omega\vert$  as it should in order to yield
the correct normalization of the delta function. The coefficient
of the second term is uniquely fixed by imposing the periodic boundary
conditions of Eq.~(\ref{bc}) (for a different, but equivalent form of the
solution see for instance Ref.~\cite{ID} p.\ 571).

The double derivative of the  Green function can be easily calculated
using the the formula
\be
{\wp}(z)=-\frac{{\rm  d}^2}{{\rm  d}z^2}\ln\Bigl(\sigma(z)\Bigr)~~,
\ee
where $\wp(z)$ is the  Weierstrass $\wp$--function. We need only  the
first few terms of its Laurent expansion about the origin
\be
\wp(z)=\frac{1}{ z^2}+\frac{\pi^4}{15}E_4(iu)z^2+\dots~~,
\label{wp}
\ee
where $E_4(\tau)$ is the second Eisenstein series. The double pole at the
origin implies that the limit for $\xi\to\xi'$ in Eq.~(\ref{limit}) is
singular.
In order to regularize this theory we keep $\xi-\xi'$ different from
zero by putting $z={\rm  e}^{i\alpha}\vert z \vert=
{\rm  e}^{i\alpha}\varepsilon/L_1$
where $\varepsilon$ is used as an ultraviolet spatial cut-off. We get
\be
Z_q^{(2l)}= 1+\frac{\varrho_o}{\sigma_o}\left[\frac{A}{8\pi^2
\varepsilon^4}+
\frac{\cos(2\alpha) c(u)}{4\pi\varepsilon^2}+
\frac{\cos(4\alpha)d(u)}{A}\right]
+\frac{\varrho_of(u)}{\sigma_oA}+O\left(\varepsilon\right)~~,
\label{reg}
\ee
with $c(u)=\left(\pi uE_2(iu)/3-1\right)$, $d(u)=\pi^2u^2E_4(iu)/60$ and
\be
f\left(u\right)=\frac{1}{2}\left\{
\left[\frac{\pi}{6} u E_2\left(i u\right)\right]^2 -
\frac{\pi}{6} u E_2\left(i u\right) + \frac{3}{4}\right\}
\label{fu}~~.
\ee
The  three terms in the square brackets of Eq.~(\ref{reg}) are cut-off
dependent quantities. They can be reabsorbed in the
renormalization
of the couplings $\sigma,\lambda$ and $\varrho$, respectively.
In fact, putting $\sigma=
\sigma_o+\delta\sigma_o$, $\lambda=\lambda_o+\delta\lambda_o$ and
$\varrho=\varrho_o+\delta\varrho_o$  in Eq.~(\ref{zeta2l}), comparison
with Eq.~(\ref{reg}) yields
\bea
\delta\sigma_o&=&-\frac{\varrho_o}{8\pi^2\varepsilon^4\sigma_o}
\label{sigma}\\
\delta\lambda_o&=&\frac{\lambda_o\varrho_o\cos(2\alpha)c(u)}
{4\pi\varepsilon^2\sigma_o}\\
\delta\varrho_o&=&\varrho_o\cos(4\alpha)\frac{d(u)}{f(u)}~~.
\label{rho}
\eea
Of course the renormalized quantities $\sigma,\lambda$ and $\varrho$
should not depend on the regularization scheme and in particular on the
choice of the parameter $\alpha$. On the other hand,
choosing $\alpha=\frac{\pi}{8}$, we get $\delta\varrho_o=0$, which shows
that  $\varrho$ is not renormalized, at least at the first perturbative
order. Thus we can safely put  the Nambu-Goto value $\varrho=1$ in the
final formula
\be
Z=\frac{\lambda}{\sqrt{u}}{\rm  e}^{-\sigma L_1L_2}
\Bigl| \eta\left(i u\right)/\eta\left(i\right) \Bigr|^{-2}
\left[1+\frac{f(u)}{\sigma L_1L_2}+O\left(\frac{1}{(\sigma L_1L_2)^2}
\right)\right]~~,
\label{2loop}
\ee
where $f(u)$ is defined in Eq.~(\ref{fu}). Note that Eq.~(\ref{2loop})
is symmetric under the exchange $L_1\leftrightarrow L_2$ as a
consequence of the functional relation
\be
E_2\left(-\frac1\tau\right)=\tau^2E_2(\tau)-i\frac{6\tau}{\pi} \, .
\label{EE}
\ee
Similarly, it can be shown that also Eqs.~(\ref{sigma}-\ref{rho})
are symmetric under such exchange.
Our two--loop result coincides with the one obtained some years ago
in the context of string theory \cite{DF}, using the $\zeta$--function 
regularization.

The fact that two completely independent regularization schemes tell us
that the coupling $\varrho$ is not renormalized suggests that this a 
general property, even if we have not yet found a rigorous argument to 
support it. Note that in our derivation the splitting 
between the cut-off dependent part and the remainder in Eq.\ (\ref{reg})
is a crucial point.
Such a  separation arises in a natural way in our regularization, 
but it is conceivable that different cut-off schemes might generate 
a different splitting. We notice, however, that the functional form of 
the two-loop contribution is preserved by the modular invariance. This 
is not obvious in our derivation because, for sake of simplicity, we 
dealt with a rectangular torus with a purely imaginary modulus 
$\tau=iu$. It is possible to develop the theory on a 
generic torus associated to an arbitrary complex modulus $\tau$. Any 
physical quantity must be invariant under the two generators of
modular group:
\bea 
S~~:~~\tau&\to-1/\tau\\
T~~:~~\tau&\to\tau+1 \, .
\eea
The terms in the square brackets of Eq.\ (\ref{reg}) are not modular 
invariant because the angle $\alpha$ has not an intrinsic geometric 
meaning. On the contrary, the function $F(iu)=f(u)$ which yields the 
functional form of the two-loop contribution (\ref{2loop}) is modular 
invariant. Its form in a general frame (i.e.\, $\tau$ arbitrary complex 
number) is given by
\be
F\left(\tau\right)=\frac{1}{2}\left\{
\left\vert\frac{\pi}{6} \Im m(\tau) E_2\left(\tau\right)-\um
\right\vert^2 +\um\right\}~~~.
\ee
Notice that Eq.~(\ref{2loop}) has no longer the
functional form of the classical approximation (\ref{class2})
even on symmetric lattices: for $u=1$ it can be easily seen 
that Eq.~(\ref{2loop}) gives
\begin{equation}
 Z~ =~ \lambda~ e^{-\sigma L^2}\left(1+\frac{1}{4\sigma L^2}\right)\,~~,
\label{zsimm}
\end{equation}
where the identity $E_2(i)=\frac{3}{\pi}$ has been used, which follows 
directly from Eq.~(\ref{EE}).

Let us stress finally that no new free parameter is introduced
within this approach.

\section{Observables and Monte Carlo Simulations}
Let us discuss the observables of the Ising model that
we can use to test the functional
form of the interface free energy.

In the finite geometry the degeneracy of
the ground state is removed: the energy of the symmetric,
$Z_2$ invariant, ground state is separated by a small
energy gap $\Delta E$ (or inverse tunneling correlation length)
from the antisymmetric ground state energy.

This energy splitting is due to tunneling
between the two vacua and is directly linked to the free energy
of the interface.
In the dilute gas approximation, in which multi--interface
configurations are summed over, but interactions between
interfaces are neglected, the energy splitting is directly
proportional to the interface partition function $Z$, where $Z$
is given by Eqs.~(\ref{1loop},\ref{class2},\ref{2loop}).
It is then easy to show (see e.g.\ \cite{SC}) that
\begin{equation}
\Delta E~=~Z(\sigma,L_1,L_2) \, ~~,
\label{Egap}
\end{equation}
where the usual factor 2 has been reabsorbed into the parameter
$\lambda$ appearing
in the definitions of $Z$.
Let us notice that $\lambda$ has the physical dimensions of an energy.
In the scaling region of the Ising model all dimensional
quantities should depend on $\beta$ according to the scaling law
\be
\xi\left(\beta\right) ~\simeq~ \xi_{\infty}
\left(1 - \frac{\beta_c}{\beta} \right)^{-\nu} \, ~~,
\label{xi}
\ee
where $\xi_{\infty}$ is the bulk correlation length in the continuum
limit. The most precise estimates for $\nu$ are in the range
from 0.624 to 0.630 \cite{betac,nuth}.
Since the interface tension is the only dimensional physical quantity
appearing in the interface free energy, in the following we
express all physical observables in units of the square root the interface 
tension $\sqrt{\sigma_{\infty}}$ according to the scaling law
\be
\sqrt{\sigma\left(\beta\right)} ~=~ \sqrt{\sigma_{\infty}}
\left(1 - \frac{\beta_c}{\beta} \right)^{\nu}~~.
\label{scal1}
\ee
The measurements of the energy gap $\Delta E$, for different choices
of the lattice sizes, provides then a first direct check on the
functional form of the interface free energy and allows one
to estimate the interface tension $\sigma$.
To this end, we have used two different methods, as explained
in the following two sections.

A second observable we have used to check our theoretical  prediction
is the surface energy $E_S$ defined by
\begin{equation}
E_S(\sigma,L_1,L_2)=-\frac{1}{Z}\frac{\partial Z}{\partial\beta}~~.
\label{ES}
\end{equation}
This observable is particularly useful because it enables us
to isolate explicitly quantum contributions beyond the Gaussian
one, as it will be discussed below.

Before describing the method we used to estimate these quantities,
let us  make some general remarks, independent from the
observable and MC method used, on the range of values of $\beta$
and on the lattice sizes where our formulae can be used.

Applying Eq.~(\ref{2loop}), one should pay attention to avoid
spurious effects like the incomplete restoration of rotational
invariance and the residual presence of lattice artifacts,
as already discussed.
To this end we made our simulation in the scaling region of
the Ising model: the lowest temperature used corresponds
to $\beta=0.240$.
(In Tab.~1 we present besides other quantities to be introduced
below estimates for the bulk correlation length of the Ising
model at the $\beta$--values used in the present study. The
MC estimate is taken from Ref.~\ \cite{XI3d}. The other estimates
are based on the low temperature series
that was extended to 15th order by Arisue \cite{Arisue}.
We analysed the series with the help of inhomogeneous
differential approximants \cite{IDA}. The numbers in the
table are the results from the $[1;7,7]$ approximation.)

The expansion parameter $\sigma A$ should be small enough to
justify the perturbative calculation.

Interactions between interfaces should be negligible, which
means that the dilute gas approximation must be satisfied: the
dominant contribution in the probability of creating an
interface is proportional to $e^{-\sigma A}$.
Too small lattice size can give rise to
a high density of interfaces and to non--negligible interactions.

Finally, the smallest size of the
lattice ($L_1$ in our conventions) should not only be greater
than the bulk correlation length but also greater than the (inverse)
deconfinement temperature of the dual gauge model.
In fact, by duality (see for instance \cite{DZ}, see also \cite{cgv}),
the broken phase of the Ising model on an asymmetric 3D lattice
corresponds to the confined phase of the 3D $Z_2$ gauge model at a
finite temperature $T^{(g)}=1/L_1$.

Precise information on the finite temperature deconfinement
transition can be found, for instance, in~\cite{wz}.
$\beta$ is mapped
to the gauge coupling constant
$\tilde\beta=-\frac{1}{2} \ln\tanh\beta$.
For each $\beta$ there exists a value $L_c(\beta)$
such that  for $L_1 \leq L_c$ the gauge system is in the deconfined
phase, and Eq.~(\ref{2loop}) cannot be applied.

The values of $L_c$ for the $\beta$'s used are given
in Tab.~1.
These have been obtained using data taken from Ref.~\cite{wz}
and the scaling law
\begin{equation}
\frac{1}{L_c(\beta)}~=~T_c^{(g)}~
(\tilde\beta_c-\tilde\beta(L_c))^{\nu}~~,
\end{equation}
where $T^c_{(g)}=2.3(1)$, $\tilde\beta_c\simeq0.7614$ and
$\nu\simeq0.630$ (see \cite{cfggv}).

\subsection{Energy Gap: the Time--Slice Correlations Method}

The first method we have used to extract the energy
splitting $\Delta E$ follows the procedure explained in
Ref.~\cite{KM} (see also Refs.~\cite{jjmmtw,cgv});
we refer to it as the {\em time--slice correlations} method
(TSC).

Consider a cylindrical geometry, with $t\gg L_1,L_2$ and periodic
boundary conditions in all directions, and define the time--slice
magnetization $S_k$,
where $k=0,1,\dots,t/2~$, as
\begin{equation}
S_k\equiv \frac{1}{L_1L_2}\sum_{n_1=1}^{L_1}\sum_{n_2=1}^{L_2}
s_{\vec n} \, ,
\end{equation}
\noindent
with $\vec n \equiv (n_1,n_2,k)$.
If one computes the two--point correlation function
\begin{equation}
G(k)\equiv\langle S_0 S_k\rangle ~~,
\end{equation}
the low energy levels of the transfer matrix spectrum can be obtained
from the asymptotic $k$--dependence of $G(k)$
\begin{eqnarray}
G(k) \cdot Z_I &=&c_0^2\left(e^{-k\Delta E}+e^{-(t-k)\Delta E}\right)+
\nonumber\\
&&c_1^2\left(e^{-k\Delta E'}+e^{-(t-k)\Delta E'}\right)
+\dots~\label{fit}\\
Z_I &=& 1 +  e^{-t \Delta E} + \dots
\end{eqnarray}
where $Z_I\equiv$ tr $e^{-t H}$ is the partition function of the Ising
model in the transfer matrix formalism.

$\Delta E'$ is the energy of the first (antisymmetric) excited state
and turns out to be (at least) one order of magnitude greater than
$\Delta E$ in the range of parameters we have used.
The coefficient $c_0$ corresponds to the magnetization expectation
value and can be used to check the consistency of the MC results.

To perform our MC simulations we used a Swendsen--Wang cluster
algorithm \cite{Wang}.
For each value of $\beta$ considered, i.e.\ $\beta=0.2246,0.2258$ and
$0.2275$, $L_1$ and $L_2$ ranged from $10$ to $35$.
We usually fixed $t=120$, using, for simulations with
particularly large $L_1$ and $L_2$, bigger sizes $t=240-360$.
The values of the energy gap $\Delta E$, extracted using Eq.~(\ref{fit}), are
reported in Tabs.~2, 3 and 4 (for Tab.~4 see also the table caption)
\footnote{Some of these data have appeared preliminarily in
Ref.~\cite{cgpv}.}; the confidence levels of these fits are always 
above $70$\%.

Using this approach one should pay attention to correlations in
MC time.
In particular it turns out that the two--point correlation
functions $G(k)$ are affected by very strong cross--correlations
in MC time.
To take under control this problem, we followed the procedure used
in Ref.~\cite{cgv},  scattering the evaluation of time--slice
correlations in Monte Carlo time.
This has the advantage of reducing the cross--correlation matrix
to an almost--diagonal form and simplify the non--linear fitting
procedure.
For each $\beta$ and each lattice we made about $0.6-1.2\cdot 10^6$
sweeps (after thermalization) with $1-2\cdot 10^3$ measurements/observable.
A standard jacknife procedure was used to evaluate errors.

Fig.~1 shows a histogram of the magnetization
for a typical lattice size.
Almost all configurations  contain zero or two interfaces, which
indicates that the dilute gas approximation is respected.
The plateau corresponds to configurations of the system
with two interfaces (because of the periodic boundary conditions in the 
z--direction). It is easy to see that the presence of more interfaces
would have the effect of strongly modify it.
The two peaks correspond to configurations
without interfaces.

\subsection{Energy Gap: the Boundary Flip Method}

Another method (which we refer to as the {\em boundary flip} (BF)
method) to evaluate the energy gap $\Delta E$ was
introduced by one of the authors in Refs.~\cite{boundary1,boundary2}.

We consider a system which allows both periodic $(p)$ and
antiperiodic $(a)$ boundary conditions $(bc)$.
The partition function of this system is given by
\begin{equation}
Z = Z_a + Z_p = \sum_{bc} \sum_{\left\{s_i=\pm 1\right\}}
e^{-\beta H\left(s , bc\right)}
\end{equation}
and the fraction of configurations with antiperiodic boundary conditions
 is given by
%\bea
%  \frac{Z_{a}}{Z}
%&=& \frac{1} {Z}\sum_{\left\{s_i=\pm 1\right\}}
%e^{-\beta H\left(s , a\right)}
% \nonumber \\
% &=& \frac{1}{Z} \sum_{bc} \sum_{\left\{s_i=\pm 1\right\}}
%\delta_{bc,a} \, e^{-\beta H\left(s , bc\right)}  \nonumber\\
%&=& <\delta_{bc,a}>~~ .
%\eea
\be
  \frac{Z_{a}}{Z}
= \frac{1} {Z}\sum_{\left\{s_i=\pm 1\right\}}
e^{-\beta H\left(s , a\right)}
 \nonumber \\
 = \frac{1}{Z} \sum_{bc} \sum_{\left\{s_i=\pm 1\right\}}
\delta_{bc,a} \, e^{-\beta H\left(s , bc\right)}  \nonumber\\
= <\delta_{bc,a}> \, .
\ee
An analogous result can be found for periodic boundary conditions.

 We can express the ratio $ {Z_{a}}/ { Z_{p}} $ as a ratio of
 observables in this system,
\begin{equation}
 \frac{Z_{a}}{Z_{p}} =\frac{<\delta_{bc,a}>}{<\delta_{bc,p}>} 
\end{equation}
which turns out to be directly connected to the energy gap $\Delta E$.

To see this, let us express the partition functions
of the periodic and antiperiodic Ising system in terms of the
transfer matrix ${\bf T}$. The antiperiodic  boundary
conditions are represented by
a spin--flip operator ${\bf P}$, which flips the sign
 of all spins in a given $z$--slice.

 The partition function of the periodic system is given by
 \begin{equation}
 Z_{p} = Tr {\bf T}^{t} ~~,
 \end{equation}
 while the partition function
 of the antiperiodic system is given by
 \begin{equation}
 Z_{a} = Tr {\bf T}^{t} {\bf P} ~~.
\end{equation}
 Since the operators ${\bf T}$ and ${\bf P}$ commute, they
 have a common set of eigenfunctions.
 Say the eigenvalues of ${\bf T}$ are $\lambda_i$ and
 those of ${\bf P}$ are
 $p_i$. The possible values of $p_i$ are $1$ and $-1$.
 States that are symmetric in the magnetization have $p_i = 1$ and
 those that are antisymmetric have $p_i = -1$.
The partition functions take the form
 \begin{equation}
 Z_{p} =  \sum_i \lambda_i^{t}
 \end{equation}
 and
 \begin{equation}
 Z_{a} =  \sum_i \lambda_i^{t} p_i ~~.
 \end{equation}
  Let us consider the ratio of the partition functions
  in the low temperature phase. If we assume that
 \begin{equation}
 \lambda_{0s} , \lambda_{0a} >> \lambda_{1s} ~, \lambda_{1a}~, ...
\end{equation}
then, ignoring terms of order
$O\left[\left(\lambda_{1s}/\lambda_{0s}\right)^t\right]$,
it is easy to show that
\begin{equation}
\left(\frac{\lambda_{0a}}{\lambda_{0s}}\right)^{t} ~=~
\frac{Z_p-Z_a}{Z_p+Z_a}~ =~ 1-2 <\delta_{bc,a}>~~.
\label{Zratio}
\end{equation}
The interface free energy (inverse of the tunneling mass) is
then given by
\begin{equation}
\Delta E~ =~ - \ln(\lambda_{0a}/\lambda_{0s})~ =~ -\frac{1}{t}
\ln\left(1-2<\delta_{bc,a}>\right)~~.
\end{equation}
Assuming that the number of interfaces is even for
periodic boundary conditions and odd for anti-periodic boundary
conditions, the dilute gas approximation leads to exactly the same
relation between the interface free energy and the boundary statistics
\cite{boundary1}.

We have used the BF method at $\beta=0.2240$ for a large  number of
lattices, as reported in Tab.~5. In a few cases we studied two or three
values of $t$, in order  to check the stability of the results. In
general, however, the $t$ value needed is   smaller than the one we
would have needed when using the TSC method. For this reason, the BF
method is particularly useful near  the critical point where large $L_1$
and $L_2$ must be  used.

For the determination of the ratio of partition
functions ${Z_a}/{Z_p}$
we employed the boundary cluster algorithm proposed by one of the
authors \cite{boundary1,boundary2}.

For each simulation given in Tab.~4 (case $c$) and in Tab.~5 we have
made, after thermalization, $0.7-1.4\cdot 10^5$ sweeps, depending
on the lattice size.

Moreover, the BF method bypasses the fitting procedure  of
Eq.~(\ref{fit}) required by the TSC method, reducing from $t/2$ to one
($<\delta_{bc,a}>$) the observables needed to evaluate $E$. This allows
to save MC time as well as drastically reduces the problems connected to
correlations in MC time.

\subsection{Surface Energy}

As stated above, one can also measure  the
surface energy and compare the MC results with the CWM
predictions.

If one assumes Eq.~(\ref{2loop}), it follows from
Eq.~(\ref{ES}) ($\sigma' \equiv \partial_{\beta}\sigma$ and
$\lambda' \equiv \partial_{\beta}\lambda$) that
\be
E_S^{(CWM)}(\sigma,L_1,L_2)=\sigma'L_1L_2 -\frac{\lambda'}{\lambda}
-\frac{1}{Z_q^{(2l)}}\partial_{\beta}Z_q^{(2l)}~~ .
\label{ES2l}
\ee
On the other hand, if one uses Eq.~(\ref{class2}) or
Eq.~(\ref{1loop}) one gets
\begin{equation}
E_S^{(1l)}(\sigma,L_1,L_2)=\sigma'L_1L_2 -\frac{\lambda'}{\lambda} \,~~,
\label{ES1l}
\end{equation}
because one--loop quantum contributions do not depend on $\beta$.
In particular, from Eq.~(\ref{2loop})
one expects corrections proportional to $(Area)^{-1}$ due to
the two--loop corrections to the CWM which would be
absent for a pure Gaussian model.

To measure the surface energy we follow the MC procedure
which was already used in Ref.~\cite{compare}.
Consider again the ratio $Z_a/Z_p$.
{}From Eq.~(\ref{Zratio}) one can express the ratio $Z_a/Z_p$
in terms of the two largest transfer matrix eigenvalues
\begin{equation}
\frac{Z_a}{Z_p} \simeq \frac{t}{2}\left(1-\frac{\lambda_{0a}}
{\lambda_{0s}}\right) \, ,
\label{Zratio2}
\end{equation}
where we again ignore terms of order
$O\left[\left(\lambda_{1s}/\lambda_{0s}\right)^t\right]$. In addition
we assume that $t << 1/\Delta E $. This means that for periodic
boundary conditions the configurations without an interface dominate while
for anti-periodic boundary conditions there is only one interface present in
almost all configurations.
Then, within this approximation, one obtains from Eq.~(\ref{Zratio2})
a definition of surface energy which does not depend on the
extension $t$, and
\begin{equation}
E_S\equiv E_p - E_a~~,
\end{equation}
with $E_p=<H>_p$ and $E_a=<H>_a$.

We have applied this method at $\beta=0.240$ for many lattice sizes
with $t$ ranging from 10 to 40. The surface energies $E_S$ for each
lattice are given in Tab.~6.

\subsection{Local Demon Algorithm }

The variance of the energy stems to a major part from fluctuations on
small scales. Hence an optimally implemented local algorithm  is superior
to a cluster algorithm in solving this particular  problem.
We used a micro-canonical demon algorithm
\cite{creutz83,creutz84,creutz86} in combination with
a particularly efficient canonical update \cite{kari93} of the demons.
This type of algorithm circumvents the frequent use of random numbers.
The algorithm is implemented using the
multi-spin coding technique \cite{creutz84,creutz86}.
Every bit of a computer word carries one Ising spin. In order to avoid
restrictions of the geometry we simulate 32 (the number
of bits in a word) independent systems.

We have chosen demons that carry the energies 4, 8, and 16.
We take one demon for each lattice site.

First we used the cycle described in \cite{compare} to simulate the
systems with periodic and anti-periodic boundary conditions
independently.  However, it turned out that  in contrast to the
situation close to the roughening transition, near the bulk
critical point also frequent updates of  the demons carrying the
energies 8 and 16 are needed to obtain autocorrelation times close to
that of the Metropolis algorithm.   As one can see from the CPU times
summarized in Tab.~1 of Ref.~\cite{compare},  this would mean a
considerable increase of the CPU time needed.

We could, however, overcome this problem by simulating systems with
periodic and  anti-periodic boundary conditions in a single simulation
and coupling them to the same demon system. It turned out that for our
particular choice of the update cycle, the integrated autocorrelation
time of $E_p-E_{a}$ was almost one order of magnitude smaller than the
integrated autocorrelation time of $E_p$ and $E_{a}$ as separate
quantities.

The cycle that we used mostly can be explained as follows. The
simulation was done by performing a cycle of 6 groups, where each group
consisted of a {\em micro-canonical} update of the 32 periodic systems
and the demon, the 32 anti-periodic systems plus the demon and    a {\em
translation} or a shift with respect to the 32 copies of the spin
system of the demon layer with energy 4, 8, or 16 (alternating). Each
group was finished by updating the demons with energy 4. The whole cycle
was completed by updating the demons with energy 8.  We performed a
measurement twice in this cycle.

For  the $24 \times 24 \times 40$ lattice at $\beta = 0.24$ we observed
the integrated autocorrelation times $\tau_p = 5.3$ and $\tau_{a} =
5.1$ of the energy with periodic and  anti-periodic boundary
conditions,   while the integrated autocorrelation time of the  surface
energy $\tau_s =  0.6$ was about one order of magnitude smaller. As
unit of the autocorrelation time we took one cycle. The simulation with
200000 cycles took about 104 h on a SPARC 10  workstation.

We used the drand48 random number generator from the C-library  and the
G05CAF from  the NAGLIB. For some choices of the parameters we simulated
with   both random number generators. The results are consistent within
the error bars.

\section{Monte Carlo Data Analysis}

In this section we compare the CWM predictions with data extracted from
MC simulations  for the two observables defined in
Eqs.~(\ref{Egap},\ref{ES}).

\subsection{FSEs of the Interface Free Energy}
Let us first discuss the
fits of the energy gap formula Eq.~(\ref{Egap})  to the data given in
Tabs.~2--5, using different functional forms  for the interface
partition function $Z$, in analogy with Refs.~\cite{PV,cgpv}.

We call {\em classical}, {\em one--loop} and {\em two--loop} fits
those made with $Z$ given by Eq.~(\ref{class2}),
Eq.~(\ref{1loop}) and Eq.~(\ref{2loop}), respectively.

The values of the energy gaps we
used range over nearly four orders of magnitude, the biggest value
being $\Delta E= 0.0444$ at $\beta=0.2246$, for a $L_1=L_2=13$ lattice
size, and the smallest $\Delta E= 0.09\cdot 10^{-4}$ at $\beta=0.2275$,
with $L_1=L_2=26$ (see Tabs.~2 and 4).

To obtain estimates of $\Delta E$ for large lattice sizes the BF
method is more efficient than the TSC method.
For this reason we have principally used it at $\beta=0.2240$
(Tab.~5), which (from our set of values) is the closest to
to the critical point.
Rather large lattices are needed here (see also Tab.~1).
The BF algorithm has been also used to take some measurements
at $\beta=0.2275$, to obtain energy gaps for particularly small
value of the two--loop contribution parameter
\be
x\equiv f(u)/(\sigma A)~~,
\label{twox}
\ee
where $f(u)$ is defined by Eq.~(\ref{fu}), with $u=L_2/L_1$ and
$A=L_1L_2$.

In Tab.~7 we report the fits we made for each $\beta$:
the reduced $\chi^2$ of the classical fits turns out to be always more
than one order of magnitude greater than those of one and two--loop fits,
in agreement with what was already shown in \cite{cgv,PV}.
Moreover, two--loop fits show a level of confidence which is
higher than the one--loop fits for all $\beta$'s: this is the first
evidence that the two--loop corrections discussed above correctly
describe the data.

For $\beta=0.2246$, $0.2258$ and $0.2275$ the MC
data\footnote{Some data at $\beta=0.2275$ and
$\beta=0.2240$ do not appear in the figures: their behaviour is
analogous to that of the other data. They have been
omitted only in order to have a better resolution in the figures.}
are given in Fig.~2, while those referring to $\beta=0.2240$, which
range over a rather different scale both in $\Delta E$ and in $A$,
are plotted in Fig.~3.
The lines represent the best two--loop fits;
their values in correspondence of the MC data are also given in
Tabs.~2--5 ($\Delta E^{(CWM)}$).

Both the high $\chi^2$ values of the fits and the clear $u$--dependence
of the MC data in Fig.~2 and Fig.~3 show that the classical
contribution, Eq.~(\ref{class2}), being only function of $A$ and not of
$u$ is completely ruled out.

Therefore, from now on, we concentrate on the one and two--loop
functional forms.

Let us now make some further considerations on the different results one
obtains with the one--loop and the two--loop approximations to the CWM.
The parameter $x$, at  fixed $\beta$ and for any given lattice, gives
the relative weight of the two--loop correction with respect to the
classical and one-loop contributions (see Eq.~(\ref{2loop})).
One expects that a fit on a sample of data
for which $x\ll 1$ should give consistent results both at one and
two--loop and that, as one increases the value of $x$, the fits made
with the
two--loop contributions should have better confidence levels
than those made with the one--loop only.
Finally, if the CWM is the correct picture,  an upper value
of $x$ should  exists at which higher order corrections
should become important and fits with the two--loop contribution
should start to show higher $\chi^2$'s.

The values of $x$ for all samples of data are given, in Tabs.~2--5
in the last columns, while the values of $\sigma$ which  have
been used to evaluate it (see Eq.~(\ref{twox})) are
given in Tab.~7 (the differences between
the one and two--loop values are not appreciable  in this case).
One can see that $x$ ranges from 0.03 up to the rather large
value of 0.27.

In order to make more stringent fits, we can use all the data we have
by expressing $\sigma$ through the scaling law (\ref{scal1}): the
free parameters then become $\lambda/\sqrt{\sigma}$,
$\sigma_\infty$ and $\nu$.
With this assumption Eqs.~(\ref{1loop},\ref{2loop}) take the form
\bea
Z_{1l} &=& \frac{\lambda}{\sqrt{\sigma}}~ \sqrt{\sigma_{\infty}}
\left(1-\beta_c/\beta\right)^\nu
 e^{-\sigma_{\infty} A\left(1-\beta_c/\beta\right)^{2\nu}}
Z_q^{(g)}\left(u\right) \label{scal1l}\\
Z_{2l}&=& Z_{1l}\cdot \left[1
+\frac{f(u)\left(1-\beta_c/\beta\right)^{-2\nu}}{\sigma_{\infty}A}+
O\left(\frac{\left(1-\beta_c/\beta\right)^{-4\nu}}{(\sigma_{\infty}A)^2}
\right)\right]~~.
\label{scal2l}
\eea
According to the usual attitude, we make the fits also using the
other scaling law
\be
\sqrt{\sigma\left(\beta\right)}~=~ \sqrt{\sigma_{\infty}}
\left(\frac{\beta}{\beta_c}-1 \right)^{\nu}
\label{scal2}
\ee
with the obvious modifications of Eqs.~(\ref{scal1l},\ref{scal2l}).

We performed the fits on the samples of energy gaps
\be
\left\{ ~~\Delta E\left(x\right) ~~| ~~x \leq x_{cut}~~ \right\}~~,
\label{sample}
\ee
starting from the first six data ($x_{cut}=0.058$) and adding
two data each time, so that the last fits contain all the
56 data.
The corresponding $\chi^2/dof$ obtained with the one--loop
(empty circles) and two--loop (full circles) formulae written
above, as functions of $x_{cut}$, are reported in Fig.~4.

{}From the figure one can see that the one--loop formula 
(\ref{scal1l}) has a $50\%$ confidence level (or more) for
$x_{cut}\sim 0.15$, which corresponds to 36 data over 56 of our
sample, while the two--loop formula (\ref{scal2l}) reaches, with
the same confidence level, $x_{cut}\sim 0.20$, which corresponds to 47 
data. The fact that both fits on the first sample have a reduced 
$\chi^2$ greater
than one is instead due to the lack of statistics (only 6 data points).
For $x_{cut}> 0.2$, the reduced $\chi^2$ of the fits made with
the two--loop correction indicates that higher order corrections
could be important. Notice, however, that the confidence levels for the 
two--loop fits are better than those for the one--loop fits.
The parameter of the fits turns out be very stable, for both types
of fits,
until the $\chi^2/dof \approx 1$: these are given in Tab.~8 for the
two limiting samples discussed above.
The results obtained for $\sigma_{\infty}$ and $\nu$ are
consistent with the values reported in the literature and, almost,
with each other, the only difference being in the constant
$\lambda/\sqrt{\sigma}$.
Let us point out that, if the two--loop contributions obtained from 
the CWM Hamiltonian (\ref{cwmham}) are physical, one expects the two
formulae (\ref{scal1l},\ref{scal2l}) to give the same constant
when the above fitting procedure is applied to a sample with
$x_{cut}\approx 0$.
As we can observe looking at our sample of data, it is rather  difficult to
reduce $x_{cut}$ below 0.05 without spoiling the reliability of the 
statistics of our analysis, 
but we can significatively check the constance of 
$\lambda/\sqrt{\sigma}$  making fits on (independent) samples in  
which the $x$ values included are approximately constant. That is, 
we define the sample
\be
\left\{ ~~\Delta E\left(x\right) ~~| ~~x_i\leq x \leq x_{i+\epsilon}~~
\right\}
\label{sample2}
\ee
and we denote with $x_{av}$ the average value of $x$ within each
sample.

The results are given in Tab.~9 and in Fig.~5.
It turns out that both the $\nu$ and $\sigma_{\infty}$ values are
very stable on the whole sample, discarding the last one where $x_{av}$
is too big.
However, while the one--loop and two--loop constants are compatible
for $x_{av}=0.04-0.07$, moving toward greater $x_{av}$ values 
$\lambda/\sqrt{\sigma}$ increases systematically in the one--loop 
case while remains stable when the two--loop formula is used.
This indicates that the absence of the $1/(\sigma A)$ corrections
in the Gaussian formula is artificially compensated by the enhancements
of its constant, already for $x_{av}\sim 0.1$.
As a final comment, let us note that a comparison with a semiclassical
$\phi^4$ approach \cite{GM} suggests to fix $\lambda/\sqrt{\sigma}$
to $\left[4~\Gamma(3/4)/\Gamma(1/4)\right]\simeq 1.352$, 
in good agreement with
the two--loop value $\simeq 1.33(1)$ of Tab.~8 and Tab.~9.

\subsection{FSEs of the Surface Energy}
In this section we discuss the finite size behaviour of the interface
partition function from MC measurements of the surface energy defined
by Eq.~(\ref{ES}).

As already said in Sec.~4.4, this approach has the advantage that
the two--loop contribution appears as an additive correction
to the surface free energy calculated at the Gaussian (and classical)
level, as given by Eq.~(\ref{ES2l}) and Eq.~(\ref{ES1l}),
respectively.
We decided to make our simulations at $\beta=0.240$ also because
very precise estimates of $\sigma$ and $\sigma'$ (two of the three
parameters on which Eqs.~(\ref{ES2l},\ref{ES1l}) depends on) are
known,
\begin{eqnarray}
      		\sigma &=&0.0590(2)~~,~{\rm from}~~\cite{HeP}
\label{s0}\\
      		\sigma'&=& 3.813(2)~~,~{\rm from}~~\cite{Ito}~~.
\label{s1}
\end{eqnarray}
Our preliminary step is to assume as external inputs these values and, 
assuming they are not biased, to use them
to test the reliability of the functional form of the second order
quantum contribution to the interface free energy.

Let us define the following surface energy differences
(at fixed $\beta$)
\begin{equation}
\Delta E_S\left(L\right)=E_S\left(2L,L/2\right)-E_S\left(L,L\right)
\label{ediff}
\end{equation}
considering two different lattices, one with $u=L_2/L_1=4$  and
the other with
$u=1$, but with the same area $A=L_1L_2=L^2$.
Then, from Eqs.~(\ref{ES1l},\ref{ES2l}) one obtains
\bea
\Delta E_S^{(1l)}\left(L\right) &=& 0
\label{DE1l}\\
\Delta E_S^{(CWM)}\left(L\right) &=&
\frac{\partial_{\beta}Z_q^{(2l)}}{Z_q^{(2l)}}\Biggl|_{u=4} -
\frac{\partial_{\beta}Z_q^{(2l)}}{Z_q^{(2l)}}\Biggl|_{u=1}~~.
\label{DE2l}
\eea
The theoretical prediction of the Gaussian model requires these
differences to be zero, while that of the CWM at two--loop does not
depend any more on the third parameter, $\lambda'/\lambda$.
Eq.~(\ref{DE2l}) can be written explicitly as
\begin{equation}
\Delta E_S^{(CWM)} = \frac{\sigma'}{\sigma^2 L^2}~
\left[f\left(4\right)-f\left(1\right)\right]
\label{ediffth}
\end{equation}
where $f(u)$ is given by Eq.~(\ref{fu}) and in the two points
we are considering takes the values $f(1)=0.25$ and
$f(4)\simeq 1.52$.

Using the MC data given in Tab.~6 (taken at $\beta=0.240$) one can
easily construct these differences.
They are given in the first column of Tab.~10, where we have assumed
$L\equiv 2L_1=L_2/2$.
It is clear that all these data are not compatible with  zero.
Assuming the validity of Eqs.~(\ref{s0},\ref{s1}) and plugging
the values into
Eq.~(\ref{ediffth}) we obtain the theoretical predictions
reported in the second column of Tab.~10. 
Monte Carlo data and 
theoretical predictions are plotted in Fig.~6

A nice $Area^{-1}$ behaviour is clearly seen in the MC data for
$L\ge16$, in good agreement with Eq.~(\ref{ediffth}).
Comparing the last column of Tab.~6 with Tab.~10 one sees that this
corresponds to $x\sim 0.1$, while for $x\geq 0.13$ higher order
corrections would be needed and for $x\geq 0.20$ the two--loop
contributions are definitively too small.

\vskip0.2cm
Then we follow the same approach of the preceding  section,
that  is we assume different forms of the
free energy and choose among them using the reduced $\chi^2$s of the
fits with equal number of parameters.
In this case, our main parameters are fixed from the fits.

We fit Eq.~(\ref{ES2l}), which can be written explicitly as
\begin{equation}
E_S^{(CWM)}\left(L_1,L_2\right)
= \sigma' L_1L_2 + \frac{\sigma'}{\sigma^2 L_1 L_2}f(u) -
\frac{\lambda'}{\lambda}
\label{enerth}
\end{equation}
and Eq.~(\ref{ES1l}).
The results are given in Tab.~11, where the parameters refer to the two--loop
fits while, for the Gaussian fits,  only the reduced $\chi^2$'s are
given in the second column.

Finally, we can make an interesting check comparing the ratio of
$\sigma'$ and $\sigma$ with the equation one obtains from the
asymptotic scaling laws.
Taking the derivative
with respect  to $\beta$ in Eq.~(\ref{scal2}), one gets
\begin{equation}
\frac{\sigma'}{2\sigma}=\frac{\nu}{\left(\beta-\beta_c\right)}~~.
\label{ratio}
\end{equation}
In order to compute this ratio we take the last two results of Tab.~11, 
the values of Eqs.~(\ref{s0},\ref{s1}) and assume $\nu\simeq 0.63$.
The results are given in the fourth column of Tab.~12.
Notice that the RHS of Eq.~(\ref{ratio}) should also be obtained
from $\lambda'/\lambda$: from Tab.~11 one sees that the two estimates
agree within errors.

On the other side, one can also assume this ratio as an input to
evaluate the critical index $\nu$, inverting Eq.~(\ref{ratio}); one
gets
\begin{equation}
\nu=\frac{\sigma' \left(\beta-\beta_c\right) }{2\sigma}~~.
\label{nu}
\end{equation}
The corresponding values are given in the last column of Tab.~12.
The agreement with the expected values is good, though the
errors are still large.

\section{Conclusions}
In this paper we have demonstrated that the finite size
effects of interface properties in the 3D Ising model
are rather well described by the two--loop expansion
of the CWM. Our MC data for the energy gap and the surface energy
prove that there are corrections to the Gaussian approximation,
and the various fits indicate that these corrections
are indeed given (up to even higher corrections) by
the two--loop result, with no extra coefficients
introduced into the the game.
Since there is (yet) no rigorous proof that the two-loop
corrections are universal (i.e., independent from the
regularization scheme) we have to consider the
predictive power of the two--loop CWM model a little
bit as a miracle.
We consider the present contribution as a first step
towards a deeper understanding of the physics of
rough interfaces in terms of effective models.

\newpage

%%%%%%%%%%%%%%%%%%%%%%%%%%%%%%%%%%%%%%%%%%%%%%%%%%%%%%%%%%%%%%%%%%
%%%%%%%%%%%%%%%%%%%%%%%%%%%%%%%%%%%%%%%%%%%%%%%%%%%%%%%%%%%%%%%%%%%
%%%%%%
%%%%%%             TABLES
%%%%%
%%%%%%%%%%%%%%%%%%%%%%%%%%%%%%%%%%%%%%%%%%%%%%%%%%%%%%%%%%%%%%%%%%%
%%%%%%%%%%%%%%%%%%%%%%%%%%%%%%%%%%%%%%%%%%%%%%%%%%%%%%%%%%%%%%%%%%%
\begin{center}
\begin{tabular}{ccccc}
\hline
      $\beta$ & $L_c$  & $L_1^{(min)}$ &
      $\xi_{\rm bulk}^{\rm MC}$ & $\xi_{\rm bulk}^{\rm IDA}$   \\
\hline
   0.2240& 12.1&  18 &         &    4.527  \\
   0.2246& 10.5&  13 &         &    3.926  \\
   0.2258&  8.5&  12 &         &    3.170  \\
   0.2275&  6.8&  10 & 2.62(2) &    2.556  \\
   0.2400&  3.4&   8 &         &    1.251  \\
 \hline
 \end{tabular}
 \end{center}
\begin{center}
{\bf Tab.~1.} {\it The first column shows the $\beta$--values
used in the present study. The corresponding inverse critical
temperature $L_c$ (in units of the lattice spacing) is
given in second column. In the third column appear the
corresponding minimal sizes used in MC simulations.
In the last two columns we present estimates for the
bulk correlation length from a Monte Carlo simulation
and from an analysis of the low temperature series
with inhomogeneous differential approximants \cite{IDA}.}
\end{center}
%%%%%%%%%%%%%%%%%%%%%%%%%%%%%%%%%%%%%%%%%%%%%%%%%%%%%%%%%%%%%%%%%%%%%%
\vskip0.2cm
\begin{center}
\begin{tabular}{ccccc}
\hline
 $L_1$ & $L_2$  & $\Delta E$   &$ ~~\Delta E^{(CWM)}$ &$x$\\
\hline
   13& 13&  0.04437(28)& 0.04441&0.226\\
   13& 26&  0.01657(28)& 0.01682&0.181\\
   13& 30&  0.01321(57)& 0.01326&0.196\\
   13& 34&  0.01069(40)& 0.01059&0.217\\
   14& 14&  0.03617(32)& 0.03627&0.267\\
   14& 28&  0.01157(62)& 0.01156&0.156\\
   14& 32&  0.00916(51)& 0.00880&0.168\\
   14& 34&  0.00723(65)& 0.00771&0.176\\
   16& 16&  0.02398(29)& 0.02355&0.149\\
   18& 18&  0.01451(46)& 0.01468&0.118\\
 \hline
 \end{tabular}
 \end{center}
\begin{center}
{\bf Tab.~2.} {\it The energy gaps $\Delta E$ obtained at $\beta=0.2246$
from MC simulations with the TSC method are reported together with
the best fit values to the CWM at 2--loop approximation given by
Eq.~(\ref{2loop}).
In the last column the values of the two--loop
parameter $x$ defined in Eq.~(\ref{twox}) are given.}
\end{center}
\newpage
%%%%%%%%%%%%%%%%%%%%%%%%%%%%%%%%%%%%%%%%%%%%%%%%%%%%%%%%%%%%%%%%%%
\vskip0.2cm
\begin{center}
\begin{tabular}{ccccc}
\hline
 $L_1$ & $L_2$  & $\Delta E$   &$ ~~\Delta E^{(CWM)}$&$x$ \\
\hline
   12& 12&  0.03750(23)& 0.03765&0.184\\
   12& 24&  0.01126(13)& 0.01117&0.212\\
   12& 26&  0.00939(13)& 0.00940&0.154\\
   12& 28&  0.00769(29)& 0.00793&0.161\\
   12& 30&  0.00665(21)& 0.00672&0.170\\
   13& 26&  0.00662(23)& 0.00684&0.126\\
   13& 28&  0.00554(32)& 0.00562&0.130\\
   14& 14&  0.02225(16)& 0.02212&0.135\\
   16& 16&  0.01236(45)& 0.01222&0.104\\
   18& 18&  0.00617(48)& 0.00631&0.082\\
 \hline
 \end{tabular}
 \end{center}
\begin{center}
{\bf Tab.~3.} {\it  The same as Tab.~2, but for $\beta=0.2258$.}
\end{center}
%%%%%%%%%%%%%%%%%%%%%%%%%%%%%%%%%%%%%%%%%%%%%%%%%%%%%%%%%%%%%%%%%%
\vskip0.2cm
\begin{center}
\begin{tabular}{ccccc}
\hline
 $L_1$ & $L_2$  & $\Delta E$   &$ ~~\Delta E^{(CWM)}$ &$x$\\
\hline
   10& 10&$  0.04334(8)^a$ &  0.04335&0.265\\
   10& 18&$  0.01439(26)^b$& 0.01457&0.131\\
   10& 20&$  0.01099(23)^b$& 0.01148&0.136\\
   10& 23&$  0.00797(23)^b$& 0.00813&0.147\\
   10& 26&$  0.00584(23)^b$& 0.00583&0.162\\
   10& 28&$  0.00396(44)^b$& 0.00469&0.174\\
   10& 32&$  0.00289(32)^b$& 0.00307&0.200\\
   11& 30&  0.00169(35)    & 0.00213&0.140\\
   11& 35&  0.00114(38)    & 0.00113&0.164\\
   12& 12&$  0.0217(1)^a$  & 0.0217&0.118\\
   12& 15&$  0.01296(21)^c$& 0.01274&0.099\\
   14& 14&$  0.00989(7)^a$ &  0.00979&0.087\\
   14& 18&$  0.00431(7)^c$ & 0.00432&0.071\\
   16& 16&$  0.00400(7)^a$ &  0.00397&0.066\\
   18& 18&$  0.00151(8)^a$ &  0.00144&0.052\\
   20& 20&$  0.000454(11)^c$ & 0.000466&0.042\\
   24& 24&$  0.000035(2)^c$ &  0.000034&0.029\\
   26& 26&$  0.000009(1)^c$ &  0.000008&0.025\\
 \hline
 \end{tabular}
 \end{center}
\begin{center}
{\bf Tab.~4.} {\it  The same as Tab.~2, but for $\beta=0.2275$.
The ($a$) and ($b$) are data taken from Ref.~\cite{KM} and
\cite{cgv}, respectively; the ($c$) data have been obtained
with the BF method.}
\end{center}
%%%%%%%%%%%%%%%%%%%%%%%%%%%%%%%%%%%%%%%%%%%%%%%%%%%%%%%%%%%%%
\vskip0.2cm
\begin{center}
\begin{tabular}{cccccc}
\hline
      $L_1$ & $L_2$& $t$  & $~~\Delta E$ &$~~\Delta E^{(CWM)}$&$x$\\
\hline
   18& 18& 54&$0.0230(2)^a$&  0.0229&0.161\\
   18& 30& 60& 0.008628(95)& 0.008650&0.124\\
   18& 35& 70& 0.005993(74)& 0.006046&0.128\\
   18& 40&120& 0.004242(42)& 0.004289&0.137\\
   18& 45& 90& 0.003034(41)& 0.003076&0.149\\
   18& 50&100& 0.002214(35)& 0.002224&0.164\\
   20& 30& 80& 0.006156(90)& 0.006149&0.101\\
   20& 35& 70& 0.004010(72)& 0.004021&0.101\\
   20& 40& 80& 0.002681(53)& 0.002669&0.105\\
   20& 45& 90& 0.001810(39)& 0.001790&0.112\\
   20& 50&150& 0.001205(14)& 0.001210&0.121\\
   24& 24& 72&$0.00657(8)^a$& 0.00646&0.091\\
   24& 35& 70& 0.001925(32)& 0.001885&0.071\\
   24& 40& 80& 0.001102(22)& 0.001107&0.070\\
   24& 45&180& 0.000652(11)& 0.000657&0.071\\
   24& 50&100& 0.000407(10)& 0.000393&0.074\\
   24& 60&120& 0.000145(5)& 0.000143&0.084\\
   30& 30& 90&$0.00136(3)^a$& 0.00133&0.058\\
 \hline
 \end{tabular}
 \end{center}
\begin{center}
{\bf Tab.~5.} {\it  Data at $\beta=0.2240$ obtained with
the BF method, where the data with index ($a$) have been taken from
Ref.~\cite{boundary1}.
The best fit values to Eq.~(\ref{2loop}) are given in
the fourth column.
The t--extensions of the lattices in the $z$--direction are also
reported.}
\end{center}
%%%%%%%%%%%%%%%%%%%%%%%%%%%%%%%%%%%%%%%%%%%%%%%%%%%%%%%%%%%%%%%%%%%%%%%%
\vskip0.2cm
\begin{center}
\begin{tabular}{cccccc}
\hline
      $L_1$ & $L_2$  &$t$ &  $E_S$   &$~~E_S^{(CWM)}$&$x$\\
\hline
       4 &16 &20    &  188.57(22) &       &0.40\\
       5 &20 &30    &  348.49(22) &       &0.26\\
       6 &24 &30,40 &  521.73(16) &       &0.18\\
       7 &28 &40    &  720.90(28) &       &0.13\\
       8 & 8 &30    &  214.19(8)  &       &0.07\\
       8 &32 &30,40 &  949.58(18) &       &0.08\\
       9 & 9 &30    &  278.85(7)  &       &0.05\\
       9 &16 &20    &  518.58(17) &       &0.04\\
       9 &36 &30,40 & 1207.47(17) &	  &0.08\\
       9 &64 &30    & 2173.99(53) &       &0.16\\
      10 &10 &30    &  350.96(7)  & 351.06&0.04\\
      10 &40 &30,40 & 1496.65(19) & 1496.47&0.06\\
      11 &11 &30    &  430.66(8)  & 430.62&0.03\\
      11 &44 &30,40 & 1816.17(25) & 1815.98&0.05\\
      12 &12 &30,40 &  517.94(9)  & 517.93&0.03\\
      12 &48 &40    & 2165.83(35) & 2166.18&0.05\\
      13 &13 &30    &  613.07(9)  & 612.95&0.02\\
      14 &14 &30,40 &  715.73(17) & 715.66&0.02\\
      15 &15 &30    &  825.83(19) & 826.04&0.02\\
      16 &16 &30,40 &  944.16(18) & 944.08&0.02\\
      18 &18 &30,40 & 1203.11(15) & 1203.12&0.01\\
      20 &20 &30,40 & 1492.86(20) & 1492.73&0.01\\
      22 &22 &30,40 & 1812.86(21) & 1812.89&0.009\\
      24 &24 &30,40 & 2163.25(23) & 2163.58&0.007\\
 \hline
 \end{tabular}
 \end{center}
\begin{center}
{\bf Tab.~6.} {\it Data for the surface energy at $\beta=0.240$
obtained by MC simulations with the local demon algorithm.
The CWM best fit values of Eq.~(\ref{2loop}) to these data are given in the
case $L_1\geq 10$.
In the last column the values of the two--loop
parameter $x$ defined in Eq.~(\ref{twox}) are given.}

\end{center}
%%%%%%%%%%%%%%%%%%%%%%%%%%%%%%%%%%%%%%%%%%%%%%%%%%%%%%%%%%%%%%%%%%%%
\newpage
\vskip0.2cm
\begin{center}
\begin{tabular}{ccccccc}
\hline
$\beta$ &   approx. &  $\chi^2/dof$& C.L.&$dof$& $\sigma$&
$\lambda/\sqrt{\sigma}$\\
\hline
0.2240&  2l& 0.63&  86\%& 16&0.004778(14)& 1.343(13)\\
      &  1l& 2.02&  1 \%& 16&0.004839(14)& 1.561(15)\\
      &  cl& 66.0&  0 \%& 16&0.004534(14)& 1.480(13)\\
0.2246&  2l& 0.54&  82\%&  8&0.006547(69)& 1.354(16)\\
      &  1l& 1.89&  6 \%&  8&0.006792(69)& 1.681(18)\\
      &  cl& 17.5&  0 \%&  8&0.005987(73)& 1.539(16)\\
0.2258&  2l& 0.45&  89\%&  8&0.009418(61)& 1.271(14)\\
      &  1l& 1.50&  15\%&  8&0.009587(61)& 1.511(15)\\
      &  cl& 20.2&  0 \%&  8&0.008363(63)& 1.324(13)\\
0.2275&  2l& 1.05&  40\%& 16&0.014728(40)& 1.332(5)\\
      &  1l& 2.36&  0 \%& 16&0.015283(40)& 1.612(6)\\
      &  cl& 18.3&  0 \%& 16&0.015114(39)& 1.594(6)\\
\hline
 \end{tabular}
 \end{center}
\begin{center}
{\bf Tab.~7.} {\it For each $\beta$, we report the best fit
results obtained using Eq.~(\ref{2loop}),  Eq.~(\ref{1loop})
and Eq.~(\ref{class2}), which correspond to 2--loop, 1--loop
and classical approximation, respectively.}
\end{center}
%%%%%%%%%%%%%%%%%%%%%%%%%%%%%%%%%%%%%%%%%%%%%%%%%%%%%%%%%%%%%%%%%%%%
\vskip0.2cm
\begin{center}
\begin{tabular}{cccccccc}
\hline
 scaling law &approx.&$\chi^2/dof$ &C.L.& $dof$&$\sigma_{\infty}$&
$\nu$&$\lambda/\sqrt{\sigma}$\\
\hline
Eq.~(\ref{scal1})&2l&0.86&74\%&44&1.47(1)&0.629(1)&1.333(8)\\
                 &1l&0.94&56\%&33&1.55(2)&0.633(1)&1.535(10)\\
Eq.~(\ref{scal2})&2l&0.88&70\%&44&1.32(1)&0.618(1)&1.331(8)\\
                 &1l&0.94&56\%&33&1.38(1)&0.622(1)&1.533(10)\\
\hline
 \end{tabular}
 \end{center}
\begin{center}
{\bf Tab.~8.} {\it  Fits on the samples defined by Eq.~(\ref{sample}),
with $x_{cut}\approx 0.2$ and $x_{cut}\approx 0.15$ for
the two--loop Eq.~(\ref{scal2l}) and the one--loop Eq.~(\ref{scal1l})
, respectively.}
\end{center}

%%%%%%%%%%%%%%%%%%%%%%%%%%%%%%%%%%%%%%%%%%%%%%%%%%%%%%%%%%%%%%%%%%%%
\newpage
\vskip0.2cm
\begin{center}
\begin{tabular}{ccccccccc}
\hline
 scaling law &$x_{av}$&approx.&$\chi^2/dof$ &C.L.& $dof$&
$\sigma_{\infty}$&$\nu$&$\lambda/\sqrt{\sigma}$\\
\hline
Eq.~(\ref{scal1})&0.04&2l&1.39&25\%&2&1.53(6)&0.636(7)&1.24(13)\\
                 &    &1l&1.50&22\%&2&1.55(6)&0.637(7)&1.33(14)\\
                 &0.07&2l&0.65&66\%&5&1.49(6)&0.630(3)&1.35(7)\\
                 &    &1l&0.72&61\%&5&1.48(6)&0.630(3)&1.43(7)\\
                 &0.10&2l&0.24&95\%&5&1.48(4)&0.629(2)&1.37(5)\\
                 &    &1l&0.35&88\%&5&1.50(5)&0.631(2)&1.48(5)\\
                 &0.12&2l&0.78&54\%&4&1.44(6)&0.627(4)&1.32(3)\\
                 &    &1l&0.82&51\%&4&1.47(6)&0.629(4)&1.49(3)\\
                 &0.14&2l&1.31&25\%&6&1.49(4)&0.630(3)&1.33(2)\\
                 &    &1l&1.96& 7\%&6&1.51(4)&0.631(3)&1.52(2)\\
                 &0.16&2l&0.78&59\%&6&1.50(7)&0.630(5)&1.35(2)\\
                 &    &1l&0.74&62\%&6&1.50(7)&0.630(5)&1.57(2)\\
                 &0.21&2l&2.87& 1\%&7&1.28(4)&0.612(3)&1.31(1)\\
                 &    &1l&6.16& 0\%&7&1.58(5)&0.634(4)&1.61(1)\\
Eq.~(\ref{scal2})&0.04&2l&1.39&25\%&2&1.37(5)&0.625(7)&1.24(13)\\
                 &    &1l&1.50&22\%&2&1.39(5)&0.626(7)&1.33(14)\\
                 &0.07&2l&0.66&65\%&5&1.33(5)&0.619(3)&1.35(7)\\
                 &    &1l&0.72&61\%&5&1.33(5)&0.619(3)&1.43(7)\\
                 &0.10&2l&0.24&95\%&5&1.33(4)&0.618(2)&1.37(5)\\
                 &    &1l&0.34&89\%&5&1.34(5)&0.620(2)&1.48(5)\\
                 &0.12&2l&0.87&48\%&4&1.29(5)&0.616(4)&1.32(3)\\
                 &    &1l&0.88&47\%&4&1.32(5)&0.618(4)&1.49(3)\\
                 &0.14&2l&1.07&38\%&6&1.34(4)&0.620(3)&1.33(2)\\
                 &    &1l&1.70&12\%&6&1.35(4)&0.621(3)&1.51(2)\\
                 &0.16&2l&0.77&59\%&6&1.35(6)&0.620(5)&1.35(2)\\
                 &    &1l&0.73&62\%&6&1.35(6)&0.620(5)&1.57(2)\\
                 &0.21&2l&2.83& 1\%&7&1.14(3)&0.600(3)&1.31(1)\\
                 &    &1l&6.19& 0\%&7&1.40(4)&0.622(4)&1.61(1)\\
\hline
 \end{tabular}
 \end{center}
\begin{center}
{\bf Tab.~9.} {\it Fits on the samples defined by Eq.~(\ref{sample2}).}
\end{center}
%%%%%%%%%%%%%%%%%%%%%%%%%%%%%%%%%%%%%%%%%%%%%%%%%%%%%%%%%%%%%%%%%%%%
\newpage
\vskip0.2cm
\begin{center}
\begin{tabular}{ccc}
\hline
       $L$   &   $\Delta E_S$    & $\Delta E_S^{CWM}$ \\
\hline
       8 &  -25.62(30)&  21.754(159)\\
      10 &   -2.47(29)&  13.923(102)\\
      12 &    3.79(25)&   9.669(71)\\
      14 &    5.17(45)&   7.103(52)\\
      16 &    5.42(36)&   5.439(40)\\
      18 &    4.36(32)&   4.297(31)\\
      20 &    3.79(39)&   3.481(25)\\
      22 &    3.31(46)&   2.877(21)\\
      24 &    2.58(58)&   2.417(18)\\
 \hline
 \end{tabular}
 \end{center}
\begin{center}
{\bf Tab.~10.} {\it Differences of surface energies at $\beta=0.240$,
obtained from Eq.~(\ref{ediff}) with the data of Tab.~6 and the
notation $L=2L_1=L_2/2$.
The theoretical predictions $\Delta E_S^{(CWM)}$ are given by
Eqs.~(\ref{ediffth},\ref{s0},\ref{s1}).}
\end{center}
%%%%%%%%%%%%%%%%%%%%%%%%%%%%%%%%%%%%%%%%%%%%%%%%%%%%%%%%%%%%%%%%%%%%
\vskip0.2cm
\begin{center}
\begin{tabular}{cccccccc}
\hline
$L_1\ge$&$\chi_{1l}^2/dof$&$\chi^2/dof$&C.L.&$dof$&$\sigma$
&$\sigma'$&$\lambda'/\lambda$ \\
\hline
  8  &107.2 & 8.02 &0.00 &17&0.0620(7) & 3.81383(24)&33.10(8) \\
  9  & 83.3 & 2.25 &0.00 &15&0.0593(8) & 3.81321(25)&33.12(9) \\
 10  & 48.5& 0.96  &0.48 &11&0.0569(12)& 3.81292(29)&33.17(13)\\
 11  & 24.1& 0.66  &0.74 &9&0.0565(18) & 3.81261(36)&33.10(17)\\
 12  & 8.4& 0.57   &0.78 &7&0.0596(37) & 3.81213(50)&32.83(27)\\
 13  & 1.7& 0.72   &0.60 &5&0.057(11)  & 3.8123(14) &32.95(87)\\
 \hline
 \end{tabular}
 \end{center}
\begin{center}
{\bf Tab.~11.} {\it Data of Tab.~6 are fitted using Eq.~(\ref{enerth}). 
For comparison, the reduced $\chi^2$'s are reported of the 
fits on the same sample of data using Eq.~(\ref{ES1l}).}
\end{center}
%%%%%%%%%%%%%%%%%%%%%%%%%%%%%%%%%%%%%%%%%%%%%%%%%%%%%%%%%%%%%%%%%%%%
\vskip0.2cm
\begin{center}
\begin{tabular}{cccccccc}
\hline
$L_1\ge$&$\sigma$&$\sigma'$&$\sigma'/(2\sigma)$&$\nu$\\
\hline
Eq.~(\ref{s0},\ref{s1})
      	   &0.0590(2) & 3.813(2)   &32.31(13) &0.593(2)  \\
       10  &0.0569(12)& 3.81292(29)&33.5(7)   &0.615(13) \\
       11  &0.0565(18)& 3.81261(36)&33.7(1.1) &0.619(20) \\
 \hline
 \end{tabular}
 \end{center}
\begin{center}
{\bf Tab.~12.} {\it The scaling ratio $\sigma'/(2\sigma)$ is evaluated
according to Eq.~(\ref{ratio}), with $\nu\simeq 0.63$ for three
estimates of $\sigma$ and $\sigma'$.
In the last column the corresponding estimates of $\nu$,
using Eq.~(\ref{nu}), are given.}
\end{center}

%%%%%%%%%%%%%%%%%%%%%%%%%%%%%%%%%%%%%%%%%%%%%%%%%%%%%%%%%%%%%%%%%%%%
\newpage
\begin{center}
{\bf Figure Captions}
\end{center}

\vskip0.5cm
{\bf Fig.~1.} {\it Histogram of the magnetization for a
typical MC ensemble at $\beta=0.2275$, with lattice sizes
$L_1=20$, $L_2=23$ and $t=120$.}
\vskip0.5cm
{\bf Fig.~2.} {\it MC data and best fit curves obtained with the
two--loop formula (\ref{2loop}) plotted for different values
of $\beta$ versus the classical interface area $A=L_1L_2$.
{}From top to bottom, the first three lines correspond to $\beta=0.2246$,
the next three to $\beta=0.2258$ and the last two to $\beta=0.2275$.
The values are also given in Tabs.~2, 3 and 4, respectively.}
\vskip0.5cm
{\bf Fig.~3.} {\it The same as Fig.~2 but for $\beta=0.2240$.}
\vskip0.5cm
{\bf Fig.~4.} {\it $\chi^2/dof$ of the fits made with the one--loop
Eq.~(\ref{scal1l}) (empty circles) and two--loop Eq.~(\ref{scal2l})
(full circles) on the samples defined by Eq.~(\ref{sample}).}
\vskip0.5cm
{\bf Fig.~5.} {\it Values of the constant $\lambda/\sqrt{\sigma}$
obtained fitting on the samples defined by Eq.~(\ref{sample2}) 
with the one--loop Eq.~(\ref{scal1l}) (empty circles) and two--loop
Eq.~(\ref{scal2l}) (full circles).}
\vskip0.5cm
{\bf Fig.~6.} {\it Comparison between Monte Carlo data and theoretical 
predictions (see Eq.~(\ref{ediffth}) in the text) for the surface 
energy differences defined in Eq.~(\ref{ediff}). The corresponding 
values are also given in Tab.~10. The dotted line is the theoretical 
expectation if two--loop corrections are neglected.}
%%%%%%%%%%%%%%%%%%%%%%%%%%%%%%%%%%%%%%%%%%%%%%%%%%%%%%%%%%%%%%%%%%%%%%

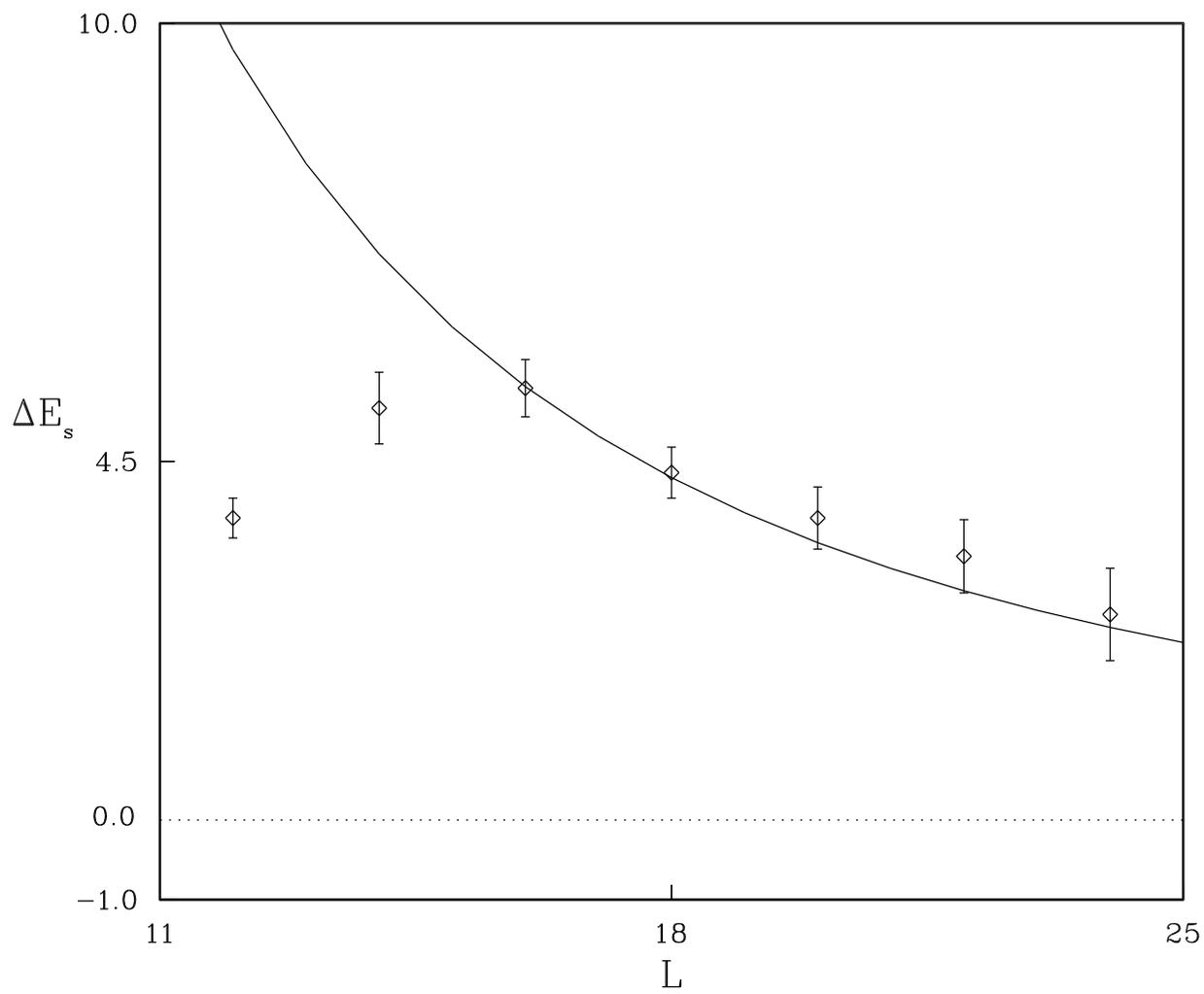
\end{document}